\documentclass[a4paper]{jpconf}
\usepackage{graphicx}
\begin{document}
\title{Unified models of the cosmological dark sector}

\author{W Zimdahl$^{1}$, W S Hip\'{o}lito-Ricaldi$^{2}$ and H E S Velten$^{1}$}

\address{$^{1}$ Universidade Federal do Esp\'{\i}rito Santo,
Departamento
de F\'{\i}sica,
Av. Fernando Ferrari, 514, Campus de Goiabeiras, CEP 29075-910,
Vit\'oria, Esp\'{\i}rito Santo, Brazil}

\address{$^{2}$ Universidade Federal do Esp\'{\i}rito Santo, Departamento de Ci\^encias Matem\'aticas e Naturais, CEUNES
Rodovia BR 101 Norte, km. 60, CEP 29932-540,
S\~ao Mateus, Esp\'{\i}rito Santo, Brazil}

\ead{winfried.zimdahl@pq.cnpq.br;hipolito@ceunes.ufes.br;velten@cce.ufes.br}

\begin{abstract}
We model the cosmological substratum by a viscous fluid that is supposed to provide a unified description of the dark sector and pressureless baryonic matter.
In the homogeneous and isotropic background the \textit{total} energy density of this mixture behaves as a generalized Chaplygin gas.
The perturbations of this energy density are intrinsically non-adiabatic and source relative entropy perturbations.
The resulting baryonic matter power spectrum is shown to be compatible with the 2dFGRS and SDSS (DR7) data.
A joint statistical analysis, using also Hubble-function and supernovae Ia data, shows that, different from other studies, there exists a maximum in the probability distribution for a negative present value $q_{0} \approx - 0.53$ of the deceleration parameter.
Moreover, different from other approaches, the unified model presented here favors a matter content that is of the order of the baryonic matter abundance suggested by big-bang nucleosynthesis.
\end{abstract}

\section{Introduction}
According to the prevailing interpretation, our Universe is dynamically dominated by a cosmological constant $\Lambda$ (or a dynamical equivalent, called dark energy (DE)) which contributes more than 70\% to the total cosmic energy budget. More than  20\% are contributed by cold dark matter (CDM) and only about 5\% are in the form of conventional, baryonic matter. Because of the cosmological constant problem in its different facets, including the coincidence problem, a great deal of work was devoted to alternative approaches in which a similar dynamics as that of the $\Lambda$CDM model is reproduced
with a time varying cosmological term, i.e., the cosmological constant is replaced by a dynamical quantity.
Both dark matter (DM) and DE manifest themselves so far only through their gravitational interaction. This provides a motivation for approaches in which DM and DE appear as different manifestation of one single
dark-sector component. The Chaplygin-gas model and its different
generalizations realize this idea. Unified models of the dark sector of this type are attractive since one and the same component behaves as pressureless matter at high redshifts and as a cosmological constant in the long time limit. While the homogeneous and isotropic background dynamics for the (generalized) Chaplygin gas (GCG) is well compatible with the data, the study of the perturbation dynamics resulted in problems which apparently ruled out all Chaplygin-gas type models except those that are observationally almost indistinguishable from the $\Lambda$CDM model. To circumvent this problem, nonadiabatic perturbations were postulated and designed in a way to make the effective sound speed vanish.
But this amounts to an ad hoc procedure which leaves open the physical origin of nonadiabatic perturbations.
There exists, however, a different type of unified models of the dark sector, namely viscous cosmological models, which are intrinsically nonadiabatic \cite{VDF}.
In the homogeneous and isotropic background a one-component viscous fluid shares the same dynamics as a GCG.  Now, what is observed in the redshift surveys is not the spectrum of the dark-matter
distribution but the baryonic matter spectrum. Including a baryon component into the perturbation dynamics for a universe with a Chaplygin-gas dark sector, there appears the new problem that the unified Chaplygin-gas scenario itself is disfavored by the data. It is only if the unified scenario with a fixed pressureless (supposedly) baryonic matter fraction of about $0.043$ (according to the results from WMAP and primordial nucleosynthesis) is \textit{imposed} on the dynamics, that consistency with the data is obtained \cite{chaprel}. If the pressureless matter fraction is left free, its best-fit value  is much larger than the baryonic fraction. In fact it becomes even close to unity, leaving only a small percentage for the Chaplygin gas, thus invalidating the entire scenario. In other words, a Chaplygin-gas-based unified model of the dark sector is difficult to reconcile with observations. One may ask now, whether the status of unified models
can again be remedied by replacing the Chaplygin gas by a viscous fluid. It is exactly this question that we are going to investigate in the present paper.
It is our purpose to study cosmological perturbations for a two-component model of baryons and
a viscous fluid, where the latter represents a one-component description of the dark sector \cite{BaVDF}.
We shall show that such type of unified model is not only consistent for a fixed fraction of the baryons but also for the case that the matter fraction is left free.  Our analysis demonstrates that the statistically preferred value for the abundance of pressureless matter is compatible with the mentioned baryon fraction $0.043$ that follows from the synthesis of light elements.

\section{The two-component model}
The cosmic medium is assumed  to be describable  by an energy-momentum
tensor $T^{ik}$ which splits into a matter part $T^{ik}_{M}$ and viscous fluid part $T^{ik}_{V} $,
\begin{equation}
T^{ik} = \rho u^{i}u^{k} + p\left(g^{ik} +
u^{i}u^{k}\right) \ , \qquad T^{ik} = T^{ik}_{M} + T^{ik}_{V}
 \, , \label{Tik}
\end{equation}
with
\begin{equation}
T^{ik}_{M} = \rho_{M} u^{i}_{M}u^{k}_{M} + p_{M}\left(g^{ik} +
u^{i}_{M}u^{k}_{M}\right) \ ,\qquad T^{ik}_{V} = \rho_{V}
u^{i}_{V}u^{k}_{V} + p_{V}\left(g^{ik} + u^{i}_{V}u^{k}_{V}\right)
\ ,\label{Tmv}
\end{equation}
where the subscript ``M" stands for matter and the subscript ``V" stands for viscous.
The total cosmic fluid is characterized by a four velocity $u^{m}$ while $u^{i}_{M}$ represents
the four velocity of the matter part and $u^{i}_{V}$ represents the four velocity of the viscous fluid.
Energy-momentum conservation is supposed to hold separately for
each of the components,
\begin{equation}
T^{ik}_{M\,;i} = T^{ik}_{V\,;i} = 0 \quad \Rightarrow\quad
T^{ik}_{\ ;i} = 0\ .\label{T;}
\end{equation}
Let the matter be pressureless, i.e. $p_{M} = 0$ and the viscous fluid be characterized by a bulk viscous pressure $p_{V} = p = - \zeta \Theta$,
where $\zeta = $ const and $\Theta = u^{i}_{;i}$ is the fluid
expansion. Under this condition the total pressure coincides with the pressure
of the viscous component. In terms of the present value $q_{0}$ of the deceleration parameter the Hubble rate  can be written as \cite{VDF}
\begin{equation}
\frac{H}{H_{0}} = \frac{1}{3}\,\left[1 - 2q_{0} + 2 \left(1 +
q_{0}\right)a^{-\frac{3}{2}}\right]\ ,
\label{r/r0q}
\end{equation}
which coincides with the Hubble rate of a specific ($\alpha = - 1/2$) GCG.
Since $\rho_{M} =
\rho_{M0}a^{-3}$, we have
$\rho_{V} = \rho -
\rho_{M0}a^{-3}$.
It is the total energy density that behaves as a GCG, not the
component $V$. This type of unified model differs from unified models in which the total energy density
is the sum of a GCG and a baryon component. Only if the baryon component is ignored, both
descriptions coincide. Consequently, in the homogeneous and isotropic background,
a generalized Chaplygin gas with $\alpha = - 1/2$ can be seen as a
unified description of the cosmic medium, consisting of a separately conserved matter component and
a bulk viscous fluid with $\zeta = $ const, where the latter itself represents a unified model of the dark sector.

\section{Perturbations}
The nonadiabaticity of the system as a whole
is characterized by
\begin{equation}
\frac{\hat{p}}{\rho + p} - \frac{\dot{p}}{\dot{\rho}}
\frac{\hat{\rho}}{\rho + p} = 3 H \frac{\dot{p}}{\dot{\rho}}
\left(\frac{\hat{\rho}}{\dot{\rho}} -
\frac{\hat{\Theta}}{\dot{\Theta}}\right)
 \, .  \label{P-}
\end{equation}
The quantity (\ref{P-}) is governed by the dynamics of the total energy-density
perturbation $\hat{\rho}$ and by the perturbations $\hat{\Theta}$
of the expansion scalar, which is also a quantity that
characterizes the system as a whole. The behavior of these
quantities is described by the energy-momentum conservation for
the entire system and by the Raychaudhuri equation, respectively. Both of these
equations are coupled to each other.
The remarkable point is that these quantities and, consequently,
the total energy density perturbation, are independent of the
two-component structure of the medium. The reason is the direct relation $\hat{p} = -
\zeta\hat{\Theta}$
between the pressure perturbations and the
perturbations of the expansion scalar.
It is convenient to describe the perturbation dynamics in terms of
gauge invariant quantities which represent perturbations on
comoving hypersurfaces, indicated by a superscript $c$. These are defined as ($v$ is the velocity potential, defined by $\hat{u}_{\mu} = v_{\mu}$)
\begin{equation}
\frac{\hat{\rho}^{c}}{\dot{\rho}} \equiv
\frac{\hat{\rho}}{\dot{\rho}} + v \ , \qquad
\delta \equiv \frac{\hat{\rho}^{c}}{\rho}
 \, .  \label{defc}
\end{equation}
Then, in linear order,
\begin{equation}
\delta'' + f\left(a\right)\delta' + g\left(a\right) \,\delta = 0 \ ,\label{dddshort}
\end{equation}
where $\delta' \equiv \frac{d \delta}{d a}$ and the coefficients $f\left(a\right)$ and $g\left(a\right)$ are given by explicitly known background quantities \cite{VDF,BaVDF}.
The relative entropy perturbations are defined by
\begin{equation}
S_{MV} \equiv \frac{\hat{\rho}_{M}}{\rho_{M}} -
\frac{\hat{\rho}_{V}}{\rho_{V} + p_{V}}  \ \label{SVM}
\end{equation}
and obey the equation
\begin{equation}
S_{VM}'' + r(a)S_{VM}' + s(a) S_{VM} = c(a) \delta' + d(a)\delta
 \ .
\label{ppS}
\end{equation}
Again, the coefficients $r\left(a\right)$,  $s\left(a\right)$,  $c\left(a\right)$ and $d\left(a\right)$
are combinations of analytically given background variables \cite{BaVDF}.
The quantity relevant for the observations is the
fractional perturbation $\delta_{M} \equiv \frac{\hat{\rho}_{M}^{c}}{\rho_{M}}$ of the energy density of the baryons, given by
\begin{equation}
\delta_{M} = \frac{1}{\gamma}\left[\delta - \frac{\rho_{V} +
p}{\rho}S_{VM}\right] \ .
 \label{del1}
\end{equation}
At early times, i.e. for small scale factors $a \ll 1$, the equation (\ref{dddshort}) has the asymptotic form
$\delta'' + \frac{3}{2a} \,\delta' - \frac{3}{2a^2}\,\delta =0 $
independent of $q_0$ and for all scales.
Eq.~(\ref{ppS}) then reduces to
$S_{VM}'' + \frac{3}{2a}S_{VM}'  = 0$ with
the solution $S_{VM} = $ const $=0$.
Consequently, there are neither nonadiabatic contributions to the total energy-density fluctuations nor
relative entropy perturbations and we have purely adiabatic perturbations $\delta_{M} = \delta$ at $a \ll 1$.
This allows us to match the initial conditions for equations (\ref{dddshort}) and (\ref{ppS}) with those of the $\Lambda$CDM model at early times.

\begin{center}
\begin{figure}[!h]
\hspace{0cm}
\begin{minipage}[t]{0.4\linewidth}
\includegraphics[width=\linewidth]{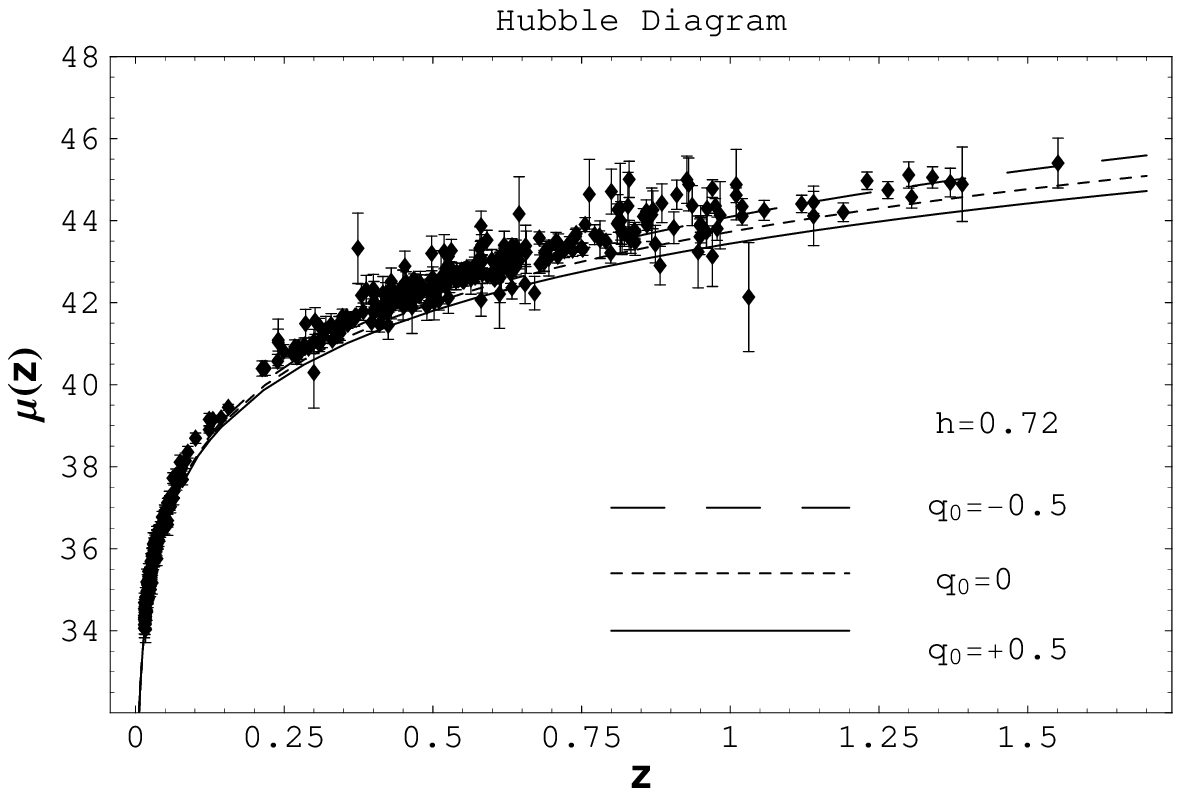}
\end{minipage} \hfill
\begin{minipage}[t]{0.25\linewidth}
\includegraphics[width=\linewidth]{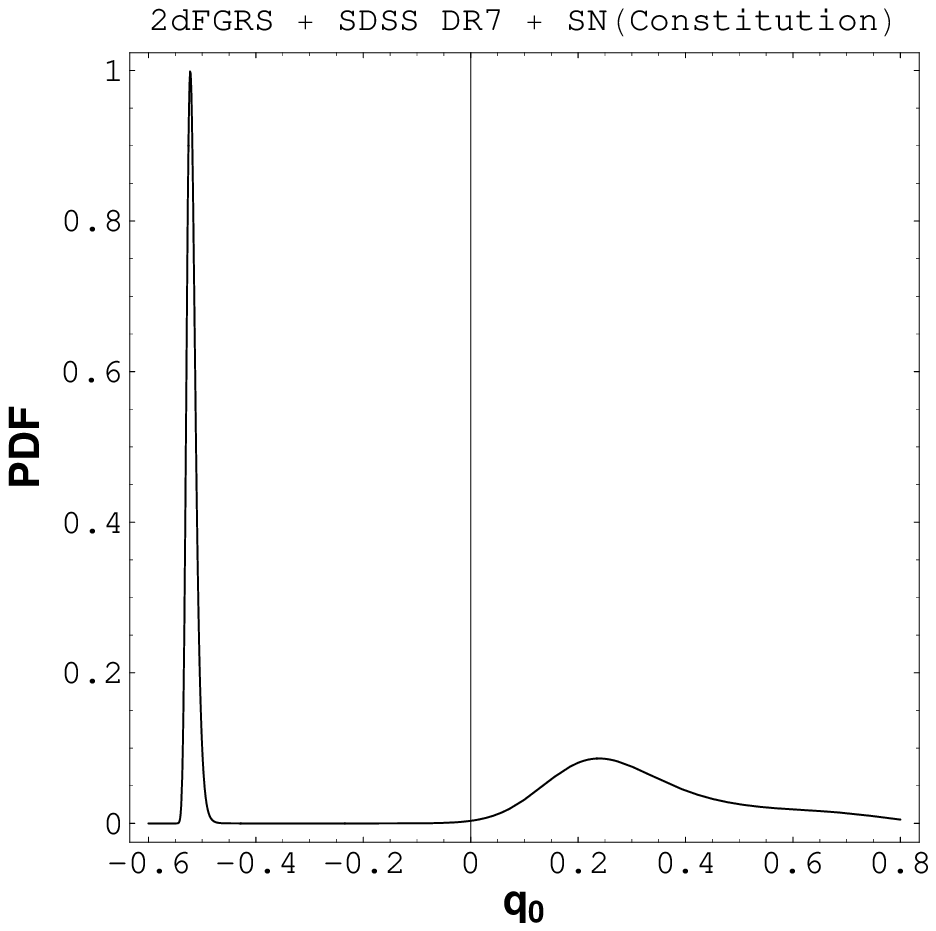}
\end{minipage} \hfill
\begin{minipage}[t]{0.26\linewidth}
\includegraphics[width=\linewidth]{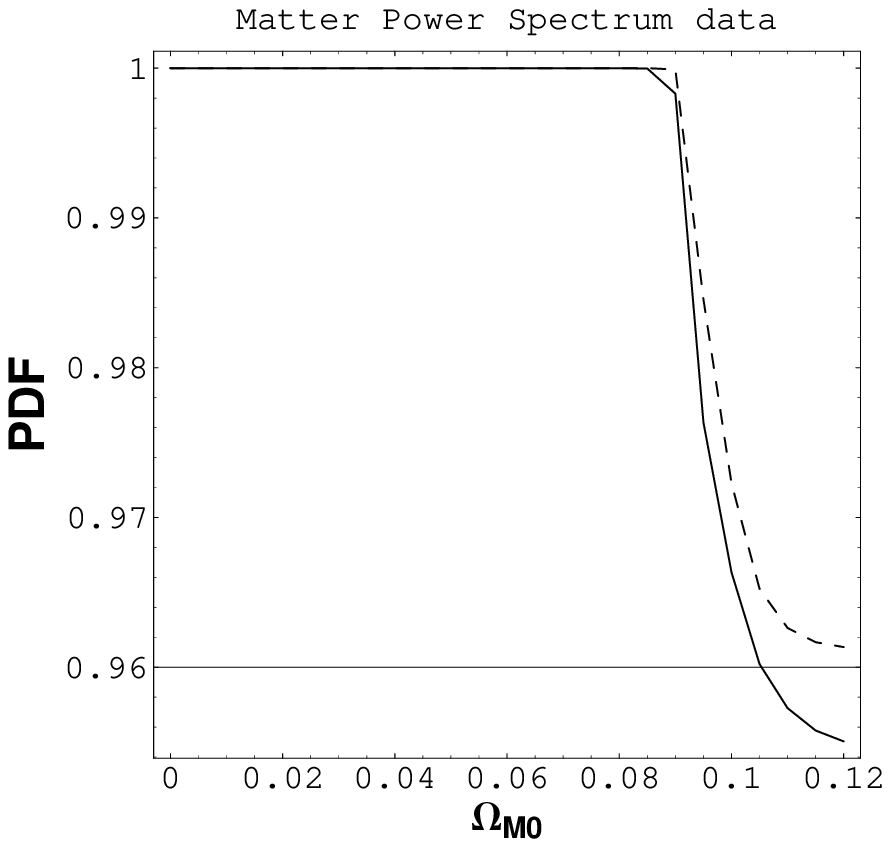}
\end{minipage} \hfill
\caption{{\protect\footnotesize Left panel: Hubble diagram, center panel:
probability distribution function (PDF) for $q_{0}$, based on a joint
analysis of matter power spectrum and SN Ia data. The right panel shows
the PDF for the pressureless component. For more details see
\cite{BaVDF}}.}
\label{SN}
\end{figure}
\end{center}

The results of our statistical analysis are shown in Fig.~\ref{SN}.


\section{Discussion and Conclusions}

The approach presented here is based on the fact that the two-component system of a bulk viscous fluid and a separately conserved baryon component behaves in the background as a generalized Chaplygin gas with
$\alpha = -\frac{1}{2}$.
While the baryon component may be considered dynamically negligible in the background, the situation is different on the perturbative level, since the observed matter agglomerations are related to baryonic density fluctuations.
These fluctuations are obtained from a combination of the nonadiabatic total energy density perturbations and relative entropy perturbations in the two-component system where the former source the latter.
The probability distribution for the deceleration parameter has a maximum at $q_{0} \approx -0.53$ which partially removes the degeneracy of previous studies which, taken at face value, were incompatible with an accelerated expansion and thus in obvious tension with results for the background.
Perhaps still more important is the test of the unified model itself. Many investigations on approaches with a unified dark sector fix the pressureless matter component to be that of the favored (by the WMAP data) baryon fraction and then check whether or not the resulting dynamics can reproduce the observations. But this does not say anything on how probable the division of the total cosmic substratum into roughly 96\% of a dark substance and roughly 4\% of pressureless matter is. To decide this question, one has to consider the pressureless matter fraction as a free parameter and to find out which abundance is actually favored by the data. Our analysis revealed that the matter fraction probability is indeed highest for values smaller than roughly 8\%. This is a result in favor of the unified viscous model. We recall that a corresponding analysis for a Chaplygin gas results in values close to unity \cite{chaprel} which seems to rule out such type of approaches.
The present viscous model, on the other hand, remains an option for a unified description of the dark sector, at least as far as the matter power spectrum is concerned.

\ack
Support by CNPq and FAPES is gratefully acknowledged.

\end{document}